# ACCELERATOR-FEASIBLE N-BODY NONLINEAR INTEGRABLE SYSTEM

V. Danilov, Oak Ridge National Laboratory, Oak Ridge, TN 37830
S. Nagaitsev, Fermi National Accelerator Laboratory, Batavia, IL 60510

*Abstract*
Nonlinear *N*-body integrable Hamiltonian systems, where *N* is an arbitrary number, attract the attention of mathematical physicists for the last several decades, following the discovery of some number of these systems. This paper presents a new integrable system, which can be realized in facilities such as particle accelerators. This feature makes it more attractive than many of the previous such systems with singular or unphysical forces.

## I. Introduction

Integrable systems with many or infinite number of degrees of freedoms attract attention due to their unique properties. For example, the Korteweg - de Vries equation is equivalent to a Hamiltonian system with infinite degrees of freedom with special waves, called solitons, particle-like scattering of them and the possibility to solve the correspondent nonlinear equations analytically [1]. A few *N*-particle systems were found for special interaction forces between particles (see, e.g., [2]). The practical realizations and applications of such systems are questionable because of very special features of such forces. Here we describe a new *N*-particle integrable system and show the practical possibilities of how to realize such a system in a particle accelerator. This system has *N* integrals of motion, which are independent and their Poisson brackets vanish.

## II. Integrable Many-Body System

Consider the following 1D mapping for the *i*-th particle, among *N* identical particles. The proposed mapping consists of two stages: a purely linear transformation and a particle interaction stage,

$$\begin{pmatrix} x_i \\ p_i \end{pmatrix}_n = \begin{pmatrix} 1 & 0 \\ K_n & 1 \end{pmatrix} \begin{pmatrix} 0 & 1 \\ -1 & 0 \end{pmatrix} \begin{pmatrix} x_i \\ p_i \end{pmatrix}_{n-1} . \quad (1)$$

One can rewrite it as

$$\begin{aligned} x_{i,n} &= p_{i,n-1} \\ p_{i,n} &= -x_{i,n-1} + K_n x_{i,n} \end{aligned}, \quad (2)$$

where $x_{i,n}$ and $p_{i,n}$ are the coordinate and the momentum of particle $i = [1...N]$ at the *n*-th mapping step. The $K_n$ term expresses the particle interaction term as follows,

$$K_n = \frac{b}{a\sigma_n^2 + 1} , \quad (3)$$

where $\sigma_n^2 = \frac{1}{N}\sum_i (x_{i,n})^2$ is the mean squared position of all particles, and $a$ and $b$ are arbitrary parameters. For $N=1$ (a single particle), the mapping (2) becomes the well-known single-particle 1D integrable McMillan mapping [3] with $\sigma_n^2 = x_n^2$. This mapping has the following integral (a conserved quantity):

$$I = ax_n^2 p_n^2 - bx_n p_n + x_n^2 + p_n^2 . \tag{4}$$

It was shown in [4] that an $N=2$ mapping is also integrable, i.e. it has two integrals of motion. One is the so-called angular momentum, $M = x_{1,n} p_{2,n} - x_{2,n} p_{1,n}$ and the second one is similar to (4):

$$I = \frac{a}{2}\left(x_{1,n} p_{1,n} + x_{2,n} p_{2,n}\right)^2 - b\left(x_{1,n} p_{1,n} + x_{2,n} p_{2,n}\right) + x_{1,n}^2 + x_{2,n}^2 + p_{1,n}^2 + p_{2,n}^2 . \tag{5}$$

We have found that the mapping (1), (2) is integrable for any $N$, i.e. it has $N$ independent integrals of motion with vanishing Poisson brackets. We first notice that all moments of the following type (for $i, j = [1...N]$ and $i \neq j$),

$$M_{i,j} = x_{i,n} p_{j,n} - x_{j,n} p_{i,n} , \tag{6}$$

are conserved. This gives $\frac{N(N-1)}{2}$ independent integrals of motion. However, only some of them have vanishing Poisson brackets. The following are $N-1$ integrals of motion with vanishing Poisson brackets:

$$I_j = \sum_{i=1}^{j} \left(M_{i,j+1}\right)^2 , \tag{7}$$

where $j = [1, 2...N-1]$. The proof can be established by induction. Assume $k-1$ invariants commute (have vanishing Poisson brackets), then, take the $k$-th integral. It consists of a sum of all terms with the new $k+1$-th variable, which is not present in all previous integrals. We omit the index $n$ for brevity and take any term from previous integrals, $(x_i p_j - p_i x_j)^2$, and the two terms from the last ($k$-th) integral $(x_i p_{k+1} - p_i x_{k+1})^2 + (x_j p_{k+1} - p_j x_{k+1})^2$ with $i, j \leq k$. It is easy to show that all other terms in this last integral have zero Poisson brackets with the first term since they do not contain any variables in common. It can also be checked algebraically that the two terms given above have zero Poisson brackets, thus proving that integrals (7) commute. The final, $N$-th integral is

$$I_N = \frac{a}{N}\left(\sum_i x_{i,n} p_{i,n}\right)^2 - b\sum_i x_{i,n} p_{i,n} + \sum_i x_{i,n}^2 + \sum_i p_{i,n}^2 . \tag{8}$$

Thus, the mapping (2) is an integrable $N$-body system.

With exception of a matched distribution (described below), which results in a perfectly linear motion, this mapping is non-linear, i.e. particle trajectories depend on the initial amplitudes of all particles through a non-linear interaction term, Eq. (3). Figure 1 shows an example of phase-space trajectories for four particles ($N = 4$).

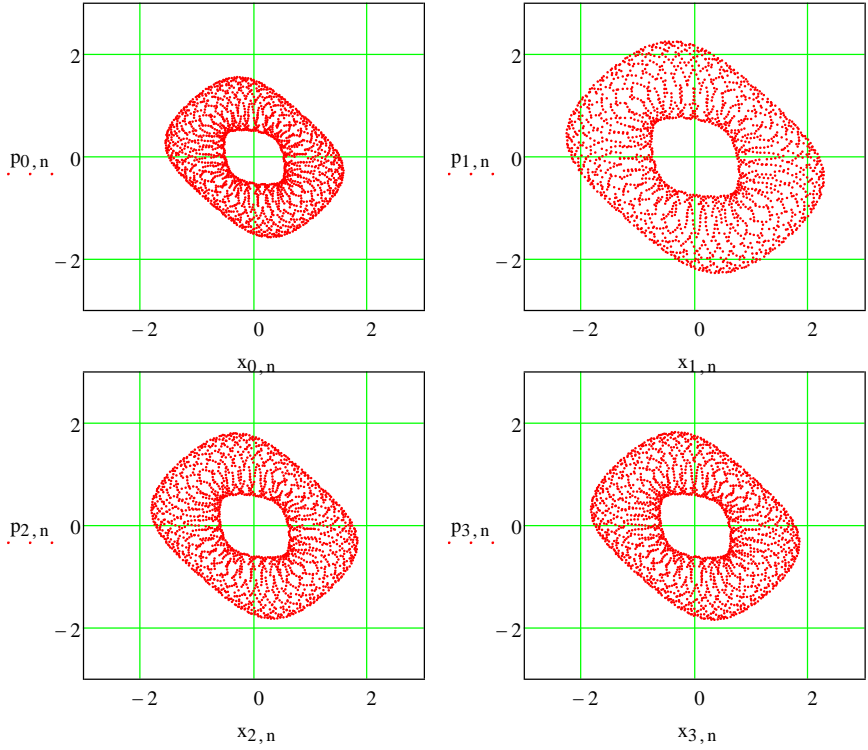

**Figure 1:** Phase-space trajectories of four particles with random initial conditions ($a = 1$, $b = -1$).

### III. Envelope Equations

Let us now describe the evolution of the $N$-particle envelope, $\sigma_n$, defined as the rms position of all particles. Following Ref. [5], we first note that the interaction term (3) acts identically on all particles and, therefore, results in a piecewise linear mapping for each step. This means that the so-called emittance,

$$\varepsilon = \sqrt{\langle x_n^2 \rangle \langle p_n^2 \rangle - \langle x_n p_n \rangle^2} \ , \tag{9}$$

is a conserved quantity ($\langle ... \rangle$ indicates averaging over $N$ particles). This also means that one can write the mapping for the envelope much like for the radial position and the radial momentum of a single particle in Ref. [4]. First, we will define the envelope momentum-like variable,

$$\eta_n = \frac{1}{\sigma_n N} \sum_i x_{i,n} p_{i,n} \ . \tag{10}$$

Now, the mapping for the envelope can be written as

$$\sigma_n = \sqrt{\eta_{n-1}^2 + \frac{\varepsilon^2}{\sigma_{n-1}^2}}$$

$$\eta_n = -\frac{\sigma_{n-1}\eta_{n-1}}{\sigma_n} + K_n \sigma_n$$

(11)

This is also an integrable mapping. The integral [4] can be written as

$$I = \left(a\sigma_n^2 + 1\right)\eta_n^2 - b\sigma_n \eta_n + \sigma_n^2 + \frac{\varepsilon^2}{\sigma_n^2} ,$$  (12)

which is actually the same integral as (8). Contrary to the rms (or envelope) equations for the Kapchinsky-Vladimirsky space charge distribution, which are chaotic in general even for the azimuthally symmetric beams [6], the equations (11) have only regular analytic solutions. Figure 2 shows an example of the envelope mapping for 4 particles with some random initial distribution (same as in Figure 1).

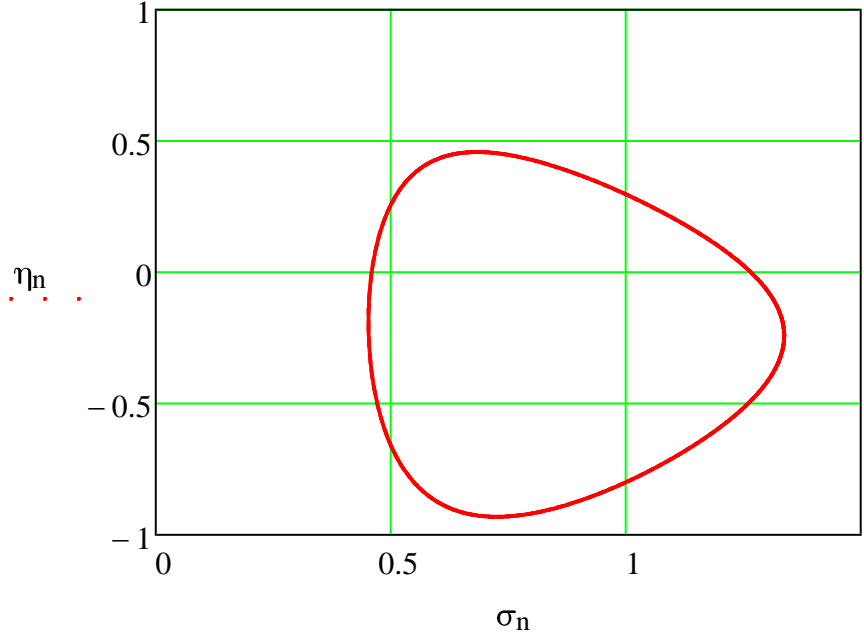

**Figure 2:** An example of the envelope mapping for 4 particles.

### IV. Matched Distribution

It is clear that for a large number of particles one can create a distribution, which maps onto itself by the mapping (2). We would call such a distribution "matched" because the interaction term, $K_n = K$, remains constant and, thus, the mapping becomes perfectly linear. Such a linear mapping preserves the so-called Courant-Snyder invariant [7],

$$J = x^2 - Kxp + p^2 .$$  (13)

Thus, it is obvious that any particle distribution function, $f(J)$, would remain unchanged under the mapping (2), provided the rms position, $\sigma = \sqrt{\int x^2 f(J)dxdp}$, satisfies the following equation:

$$K = \frac{b}{a\sigma^2 + 1} . \qquad (14)$$

## V. Practical Realization

Finally, let us describe how to realize such a mapping in a particle accelerator. The beam dynamics, described by (2), is fairly straightforward to realize in a circular accelerator. One needs to measure the beam r.m.s. position, $\sigma$, which is a variable in equations (2) and apply an integral quadrupole kick with the coefficient $\frac{b}{a\sigma^2 + 1}$ to the circulating beam, where $\sigma$ of the beam has to be taken at the quadrupole position. It can be achieved, for example, by placing the profile measurement device and the quadrupole such that the betatron phase advance between them is an integer of π. The quadrupole kicks themselves have to be separated by π/2 betatron phase advance. This is in a sense a standard feedback, but the beam parameter of interest is its rms position, not the displacement.

## Summary

An example of a rare *N*-body integrable nonlinear system was found. Its unique properties allow one to create such a system experimentally in accelerators, opening a new venue for various applications to charged particle beams.

## Acknowledgements

This research is supported by UT-Battelle, LLC and by FRA, LLC for the U. S. Department of Energy under contracts No. DE-AC05-00OR22725 and DE-AC02-07CH11359 respectively.